\documentclass[doublecol]{epl2}

\title{New insights into black bodies}
\shorttitle{New insights into black bodies} 

\author{F. J. Ballesteros\thanks{E-mail: \email{fernando.ballesteros@uv.es}}
}
\shortauthor{F. J. Ballesteros}

\institute{
  Astronomical Observatory of the University of Valencia (OAUV) - Ed. Instituts d'Investigaci\'o, \\Parc Cient\'ific, C/ Catedr\'atico Jos\'e Beltr\'an 2. E-46980 Paterna, Valencia, Spain
}

\abstract{Planck's law describes the radiation of black bodies. The study of its properties is of special interest, as
   black bodies are a good description for the behavior of many phenomena.
   In this work a new mathematical study of Planck's law is performed and new properties of this old acquaintance
   are obtained. As a result, the exact form for the locus in a color-color diagrams has been deduced, and
   an analytical formula to determine with precision the black body temperature of an object
   from any pair of measurements has been developed. Thus, using two images of the same field obtained with
   different filters, one can compute a fast estimation of black body temperatures for every pixel in the image, that is,
   a new image of the black body temperatures for all the objects in the field. Once these temperatures are obtained,
   the method allows, as a consequence, a quick estimation of their emission in other frequencies, assuming a black body behavior.
   These results provide new tools for data analysis.}

\begin{document}

\maketitle

\section{Introduction}
\label{intro}

   Many interesting phenomena emitting thermal radiation follow approximately the theoretical
   curve of a black body: hot stars, volcanic lava,
   the cosmic microwave background or the infrared emission from the Earth, for instance.
   The determination of their temperature and other properties is a frequent topic,
   and Planck's law becomes a source of information about them.

   Planck's law gives the spectral distribution of electromagnetic
   emission for a black body at a given temperature. When the black body is much
   bigger than the detection element field of view, obtaining its temperature from a measurement is trivial:
   although the intensity arriving to the detector from an area unit decreases inversely with the
   square of the distance, the amount of area observed increases proportionally with the
   square of the distance, and both effects compensate each other (as in
   infrared land-surface temperature measurement from space, \cite{ref1}).

   When this is not the case, the received flux changes inversely with the
   square of the distance, hence the distance to the object (and its size) matters.
   If this distance is unknown, it should be enough to obtain the ratio of the intensities measured
   at two different wavelengths, as it is independent of the
   distance and size of the emitter and determines unambiguously the temperature. But then, what one get is a transcendental equation: it
   is not possible to find analytically the value of the temperature from this ratio. One has to use
   numerical methods to solve the equation for T \cite{ref2,ref3}, or fit measurements
   into a Planck's curve by means of a minimization procedure \cite{ref4} (or if the maximum of emission can be
   determined taking more points, Wien's law can also be used).

   In this paper we show a new mathematical study on Planck's
   law for black body radiation \cite{ref5}, finding some new properties. As a result
   we have deduced the general form of the stellar locus straight region in a color-color diagram (where stars with a behavior
   closer to a black body are). We have also developed a new, fast and accurate analytical formula for obtaining
   the temperature of a black body from this intensity ratio measured at two wavelengths. This method
   allows one to produce thermal images of objects taken through two filters, and as a consequence to generate
   synthetic images of these objects at any other wavelength, following a black body behavior.

\section{Color-color diagram as a power law}

In a color-color diagram, many stars fall into a rather linear stellar locus.
This is because stars approximate blackbody radiators. But although it is well known that black bodies in a color-color
diagram align into a straight line, the exact form of the formula they follow has never been deduced until now.
A color index is the difference in magnitude in two bands, which corresponds to
the ratio of the respective intensities, as the magnitude is a logarithmic function of the intensity. Hence,
a linear locus is equivalent to a power law in the space of intensity ratios.

Planck's law has the general form:
   \begin{equation}
      I = C \frac{x^n}{{e^{x/T}  - 1}}
   \end{equation}
where $C$ is a constant, $ x = h\nu /k = hc/k\lambda $ (being $\nu$ the
frequency and $\lambda$ the wavelength), $T$ is the temperature of the black body, and $n$ is an integer whose
value is linked to the flux units used, according to the following rules:

\begin{tabular}{@{}lcccccc}
  \hline
  intensity units (proportional to) & n\\
  \hline
  photons/frequency unit  & 2\\
  energy/frequency unit & 3\\
  photons/wavelength unit & 4\\
  energy/wavelength unit & 5\\
  \hline
 \end{tabular}

When we represent $I_a/I_b$ vs. $I_c/I_d$ for black bodies with different temperatures
(where $I_i$ is the intensity at wavelength/frequency
$i$ at a given temperature $T$), these intensity ratios follow a power law such as:
\begin{equation}
I_c/I_d = A (I_a/I_b)^\alpha
\end{equation}
where $A$ and $\alpha$ are coefficients to be determined. First, to obtain $\alpha$ we will use a low temperature approximation,
as $\alpha$ is mainly determined by this regime.
Substituting eq. (1) into (2) and applying logarithms, $\alpha$ gives:
\begin{equation}
\alpha  = \frac{{{\mathop{\rm Ln}\nolimits} \left( {\frac{{x_c^n }}{{x_d^n }}\frac{{e^{x_d /T}  - 1}}{{e^{x_c /T}  - 1}}} \right) - {\mathop{\rm Ln}\nolimits} (A)}}{{{\mathop{\rm Ln}\nolimits} \left( {\frac{{x_a^n }}{{x_b^n }}\frac{{e^{x_b /T}  - 1}}{{e^{x_a /T}  - 1}}} \right)}}
\end{equation}

In the low temperatures limit, $T \downarrow  \downarrow $, then  $(x/T) \uparrow \uparrow$, thus $e^{x/T}  - 1 \approx e^{x/T} $. Substituting,
and taking into account that the terms of the form $x_i/T$ dominate, we obtain the exponent of the power law, i.e. the slope for
a color-color diagram $c-d$ vs $a-b$ for black bodies. Note that it only depends on the frequencies/wavelengths considered:
\begin{equation}
\begin{array}{r}
\alpha  \approx \frac{{{\mathop{\rm Ln}\nolimits} \left( {\frac{{x_c^n }}{{x_d^n }}} \right) + \frac{{x_d }}{T} - \frac{{x_c }}{T} - {\mathop{\rm Ln}\nolimits} (A)}}{{{\mathop{\rm Ln}\nolimits} \left( {\frac{{x_a^n }}{{x_b^n }}} \right) + \frac{{x_b }}{T} - \frac{{x_a }}{T}}}
\approx \frac{{\frac{{x_d }}{T} - \frac{{x_c }}{T}}}{{\frac{{x_b }}{T} - \frac{{x_a }}{T}}} =\\= \frac{{\nu _c  - \nu _d }}{{\nu _a  - \nu _b }} = \frac{{\frac{1}{{\lambda _c }} - \frac{1}{{\lambda _d }}}}{{\frac{1}{{\lambda _a }} - \frac{1}{{\lambda _b }}}}
 \end{array}
\end{equation}

For $A$, let us consider the high temperatures limit, as values in this regime tend to crowd together,
reaching a fixed point when $T \to \infty$. Now $(x/T) \downarrow \downarrow$,
and thus $e^{x/T} \approx 1 + x/T $. Therefore the intensity ratios become:
\begin{equation}
\frac{{I_c }}{{I_d }} = \frac{{x_c^n }}{{x_d^n }}\frac{{\frac{{x_d }}{T}}}{{\frac{{x_c }}{T}}} = \frac{{x_c^{n - 1} }}{{x_d^{n - 1} }}
\hspace{20pt} \mbox{   }
\frac{{I_a }}{{I_b }} = \frac{{x_a^{n - 1} }}{{x_b^{n - 1} }}
\end{equation}

Resolving eq. (2) for $A$ it gives:
\begin{equation}
A = \frac{{I_c }}{{I_d }}\left( {\frac{{I_b }}{{I_a }}} \right)^\alpha   = \frac{{x_c^{n - 1} }}{{x_d^{n - 1} }}\left( {\frac{{x_b^{n - 1} }}{{x_a^{n - 1} }}} \right)^{\frac{{x_c  - x_d }}{{x_a  - x_b }}}
\end{equation}

Substituting (6) and (4) in eq. (2) we obtain the power law we were looking for, which is a good fit of data from black bodies. Note that we
can interrelate black body intensities in only three spectral positions, just by choosing $d=b$, and so $I_d=I_b$, $x_d=x_b$
(a smaller reduction cannot be afforded as in that case we would only obtain a trivial identity). Hence, $I_c$
can be written as:
\begin{equation}
I_c  = \left[ {\left( {\frac{{x_c }}{{x_b }}} \right)^{n - 1} \left( {\frac{{x_b }}{{x_a }}} \right)^{(n - 1)\frac{{x_c  - x_b }}{{x_a  - x_b }}} } \right]I_b \left( {\frac{{I_a }}{{I_b }}} \right)^{\frac{{x_c  - x_b }}{{x_a  - x_b }}}
\end{equation}

\section{Data checking}

\begin{figure}
 \includegraphics[width=\columnwidth]{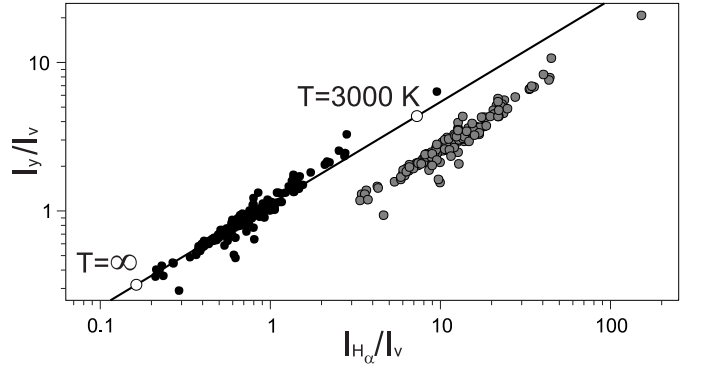}
 \caption{ Dots: data from stars in NGC 6910, taken with University of Valencia's telescope TROBAR with Stromgren filters;
gray and black dots stand respectively for data before and after correction of atmospheric extinction, exposure times and filter band width. Solid line: theoretical curve
from eq. (7) for black bodies, where two points have been stressed (white dots) at $T=3000$ K and $T=\infty$.}
\label{fit}
\end{figure}

As eq. (7) has been obtained using several approximations, its accuracy is relative: it is good
for interpolations (when spectral position $c$ lies in between $a$ and $b$) but not so good for extrapolations.
Nevertheless, it is quick to compute, rather compact and keeps the main features of the power law.
Thus it can be useful as is, taking into account its limitations, to check if
astronomical data have been properly reduced (provided that there are enough stars in the images with a good black body behavior).
In the example in fig. ~\ref{fit}, gray dots are data from the stars in the open cluster NGC 6910,
once the images were only corrected of additive components and flat-fielded, and black points are the same data
after correcting other multiplicative factors  as atmospheric extinction, exposure times and filter band width.
Solid line is the theoretical curve from eq. (7), which fits perfectly the black dots, showing that the data reduction has been correctly done.

Although similar in philosophy to methods that fit stars to a stellar locus,  as
the Stellar Locus Regression method \cite{ref6}, it is very different. SLR is a method for calibration,
giving magnitudes as output, but only applicable to a given set of filters for which a standard experimental stellar locus has
been defined. On the contrary, our application is completely general, valid in principle for any set of filters. It only needs that
some of the stars behave more or less as black bodies (which is frequent). It is good for checking data reduction, but as
eq. (7) deals with flux ratios, magnitude information cannot be recovered. Any position of the black dots along the line in fig. ~\ref{fit}
could work (it will be equivalent to changing the temperature of the stars). There is in principle no way to discriminate among all the possibilities. Nevertheless, some limits can be
imposed, because there is an absolute upper limit, eq. (5), corresponding to the asymptotic limit for high temperatures (not black body beyond
this point is allowed), and one can also set by hand a lower limit (as for example, $T=3000$ K, as M stars differ rather from a black body).

Although it cannot be directly used to calibrate data, if we manage to have data well calibrated for two given filters by other means, the method could indeed be used for calibrate data in
other filters, as $\alpha$ depends only on the effective wavelength. This can be useful for multiband surveys. The method can also be used to measure the effective wavelength midpoint of a unknown filter by using
two well known filters. This can be useful to recover information on the filters used
for the case of old data, when logs are not conserved and filter information has been lost.

\begin{figure}
 \includegraphics[width=\linewidth]{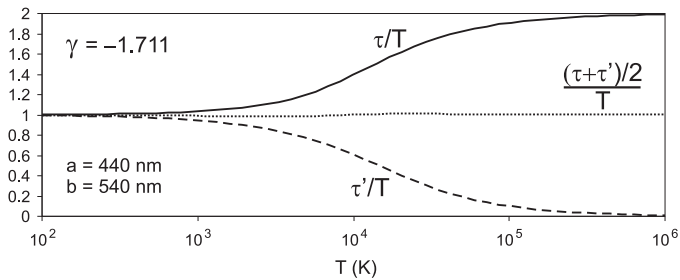}
 \caption{Solid line: ratio between the magnitude $\tau$ of eq. (9), calculated from the intensities of several black bodies
at wavelengths 440 and 540, and the real temperature of these black bodies. Dashed line: the same but for $\tau'$ following eq. (11).
Dotted line: the average of both magnitudes, eq. (12).} \label{tau}
\end{figure}

\section{The temperature of a black body}

For the purpose of obtaining directly the temperature of a black body from an intensity ratio, we take
eq. (7) as starting point. There, given two fixed spectral positions $a$ and $b$, $I_c$ varies only with $x_c$. Therefore, eq. (7) can be rewritten as:
\begin{equation}
I_c  = C'\frac{{x_c^{m} }}{{e^{x_c/ \tau} }}
\end{equation}
with
\begin{equation}
\tau = \frac{{x_b  - x_a }}{{{\mathop{\rm Ln}\nolimits} \left[ \frac{I_a }{I_b } {\left( \frac{x_b }{x_a } \right)^{n - 1} } \right]}}
\end{equation}
and
\begin{equation}
C' = I_b \left( {\frac{1}{{x_b }}} \right)^{n - 1} \left( {\frac{{x_b }}{{x_a }}} \right)^{\left( {n - 1} \right)\frac{{ - x_b }}{{x_a  - x_b }}} \left( {\frac{{I_a }}{{I_b }}} \right)^{\frac{{ - x_b }}{{x_a  - x_b }}}
\end{equation}
where $m$ is an integer.
Except for the lack of a "$-1$" term in the denominator, the formula for $I_c$ is analogue to the one of a black body, eq. (1).
Thus it is tempting to identify $\tau$ with a temperature (it has temperature units too).

Indeed, the behavior of $\tau$ is surprising.
In the low temperatures regime (when the black body's maximum $x_{\rm max}$ $< x_a,x_b$), $\tau$ estimates
correctly the temperature, $\tau \approx T$, but in the high temperatures regime ($x_{\rm max}$ $> x_a,x_b$),
it tends to  $\tau \approx 2T$, as can be seen in fig. ~\ref{tau}, solid line, where we have used data from black bodies at many
temperatures, estimating $\tau$ from their intensities at wavelengths 440 and 540 nm as an example (although
the same behavior can be find with any other pair of spectral positions).

Taking this property into account, we are going to define a new estimator for the temperature of a black body. The trick will be
to average $\tau$ with a modification of itself (namely $\tau'$) that will coincide with $\tau$ for the
low temperature regime and that will be negligible respect to $\tau$ for the high temperature regime. This can be afforded by
changing the exponent $n-1$ by $n-\gamma$ (where $\gamma$ is a numerical value that has to be fitted to assure
such behavior, and that depends on the spectral positions $a$ and $b$ used):
\begin{equation}
\tau' = \frac{{x_b  - x_a }}{{{\mathop{\rm Ln}\nolimits} \left[ \frac{I_a }{I_b } {\left( \frac{x_b }{x_a } \right)^{n - \gamma} } \right]}}
\end{equation}
For 440 and 540 nm, a good choice of $\gamma$ is $-1.711$ (see
fig. ~\ref{tau}, dashed line).
Then, our temperature estimator $\tilde T$ will be the average of both, $\tilde T = (\tau + \tau')/2$:
\begin{equation}
\begin{array}{l}
\tilde T  = \frac{{x_b  - x_a }}{{2}}  \left[ {
\frac{{1 }}{{{\mathop{\rm Ln}\nolimits} \left[ \frac{I_a }{I_b } {\left( \frac{x_b }{x_a } \right)^{n - 1} } \right]}} +
\frac{{1 }}{{{\mathop{\rm Ln}\nolimits} \left[ \frac{I_a }{I_b } {\left( \frac{x_b }{x_a } \right)^{n - \gamma} } \right]}}
}\right]\\
\end{array}
\end{equation}

\begin{figure*}
 \includegraphics[width=\linewidth]{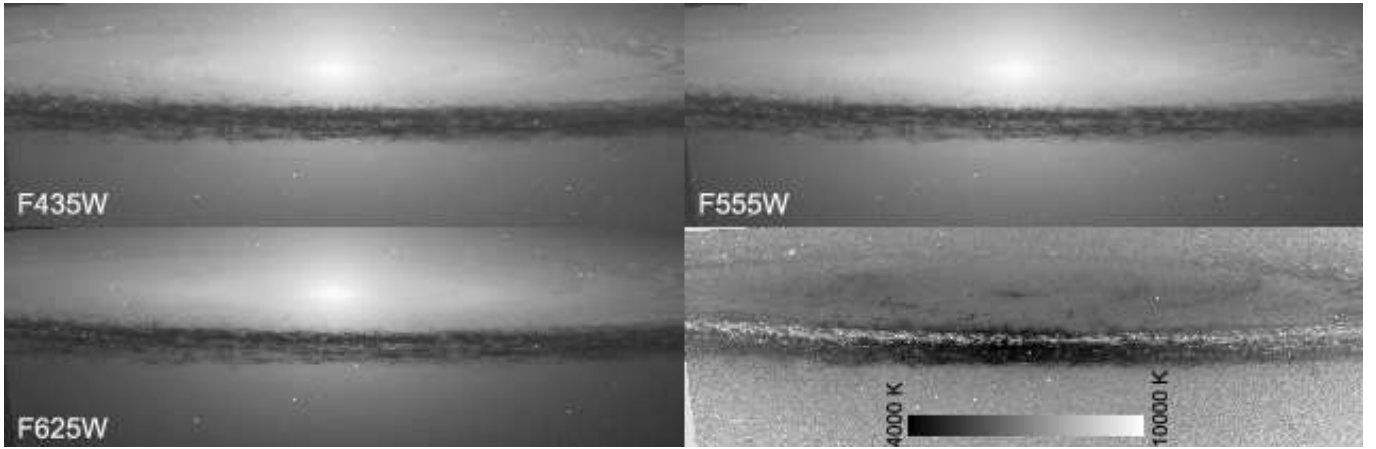}
 \caption{Images of the Sombrero Galaxy taken by the ACS WFC instrument
on board NASA/HST, with the filters F435W,
F555W and F625W respectively (the images are not reproduced here in linear scale in order to
show all the richness of details they have), and the image of equivalent
black body temperatures for the galaxy (bottom right, in linear scale), obtained
assuming a black body behavior for each pixel (black means a "temperature"
smaller than or equal to 4000 K; white, greater than or equal to 10000K). Note that this does
not afford for a real temperature of the dust; it seems that this color is produced by
unresolved stars in the foreground, which give a bigger blue component to the light from the dust.} \label{thermal}
\end{figure*}

In the low temperature regime, $\tau' \approx \tau$, thus, $\tilde T \approx 2\tau/2 = \tau \approx T$. In the high temperature regime, $\tau' << \tau$, thus, $\tilde T \approx \tau/2 \approx T$. As can be seen in fig. ~\ref{tau}, dotted line, the discrepancy with the real $T$ is minimal.

Note that for two given spectral positions $a$ and $b$, $\gamma$ has to be carefully chosen in order to ensure the best results, as $\gamma$ is a function of $x_a$ and $x_b$ (a function that has to be invariant under the interchange $a \leftrightarrow b$). Now the art lies in choosing a suitable function $\gamma$ which gives good results. Among the studied candidates it has been found that the function which works best in the whole spectrum is:
\begin{equation}
\gamma = 5 - 9[|{\mathop{\rm Ln}\nolimits} (x_a/x_b)| + 3]^{-0.252}
\end{equation}

Of course, other possible choices of $\gamma$ deserve consideration. But with this one,
the estimation of black body temperature from $I_a/I_b$ is
very precise in the whole spectral range, with a precision under $1 \%$ for most cases (the bigger error
happens when $x_{\rm max}$ lies in between $x_a$ and $x_b$; here in some cases, the error can rise
up to $\sim 5 \%$). Although in many inverse problems one finds often unexpectedly strong numerical instabilities,
this is not the case. The solution is a well-behaved formula, very robust. As an example, although $\gamma$ of eq. (13) was
in principle designed to assure a good behavior in the visible spectrum, let us use data coming for the Cosmic Microwave Background
spectrum measured by FIRAS on board COBE \cite{ref7}:

at 10 cm$^{-1}$, $I = 200$ MJy/sr.

at 20 cm$^{-1}$, $I = 10$ MJy/sr.

From this data, using eqs. (12) and (13) we get for the CMB a temperature estimation of $\tilde T=2.72$ K.

Statistical fluctuation can affect the estimation, being
more sensible to it when $x_a$ and $x_b$ are closer or when the black body temperature
is higher. For example, a SNR of 10 in $I_a$ and $I_b$, becomes a SNR in the estimated
black body temperature of: 10 for a 6000 K black body measured at 440 and 640 nm; 6 for
a 9000 K black body at the same filters; 6 for a 6000 K black body measured at 440 and 540 nm.
But note that this is not a problem of the estimator but of the black body curve itself.

All the former relationships were deduced for monochromatic emission.
When observing black bodies through filters, these relationships are right for narrow
filters (once dividing the intensities by the bandwidth), but for very broad filters as Johnson-Morgan ones,
the linear behavior in the color-color space
becomes slightly curved (as the effective wavelength of the filter changes with the shape
of the black body). Nevertheless, they still work rather well.

As a direct application of eq. (12), it can be use to estimate the temperature from
the widely used color index $B-V$.  Equation (13) is a generic fit
which works well in the whole spectrum and makes for you the selection of a $\gamma$. But for a particular
case, one can find a value of $\gamma$ that works still better than the one
produced by eq. (13). After this fine tuning, the final result is:
\begin{equation}
\begin{array}{l}
T = 4600\left( {\frac{1}{{0.92(B - V) +1.7}} + \frac{1}{{ 0.92(B - V)+0.62}}} \right)\\
\end{array}
\end{equation}
comparable to other results in the literature
(for a comparison, see \cite{ref8} where $B-V$ vs. $T$ data
for different stars are represented together with several fittings).

\section{Thermal imaging}

Equations (12) and (13) provide a tool to obtain quickly,
from two images of the same part of the sky taken through two filters,
a new image of equivalent black body temperatures, once the images have been properly flat-fielded, calibrated, etc. (such that
the same gray tone on both images is proportional to the same flux intensity) and aligned.
As eqs. (12) and (13) are analytical formulae, their calculation is fast, so it can be done for every
pixel. This way we get a new image, where gray tones are proportional to the
temperature that each pixel should have if fluxes were coming from a black body.

As an example, this method has been applied to the images of the Sombrero galaxy obtained
by NASA Hubble Space Telescope, with the ACS WFC instrument. Concretely the filters F435W,
F555W and F625W have been used, see fig. ~\ref{thermal}.

To obtain an estimation of temperatures, we just need two filters, thus three possible combinations
are available: a thermal image using F435W and F555W, another with F435W and F625W,
and still another with F555W and F625W. Of course, the sky areas corresponding to
these pixels are not real black bodies, and they
have statistical fluctuations too, so we will not get the same temperature estimations in the three cases.
But nevertheless the three images obtained are coherent: temperatures are indeed quite similar and keep a good
agreement, showing that the method is robust. In fig. ~\ref{thermal},
the bottom right image corresponds to the averaged image of these three.

In general this image shows what one should expect: dust features have an equivalent black
body temperature lower than the rest of the galaxy, which means that light from these zones is reddish,
due to the reddening caused by the dust. In Earthward direction, light
from the center of the galaxy crosses the dust ring directly towards us,
suffering a reddening more intense than the one coming from the rest of the dust ring, where disperse (and hence bluer) light
contribution is higher. And in fact, the frontal zone of the dust ring in the "thermal" image is darker.

\begin{figure}
 \includegraphics[width=\columnwidth]{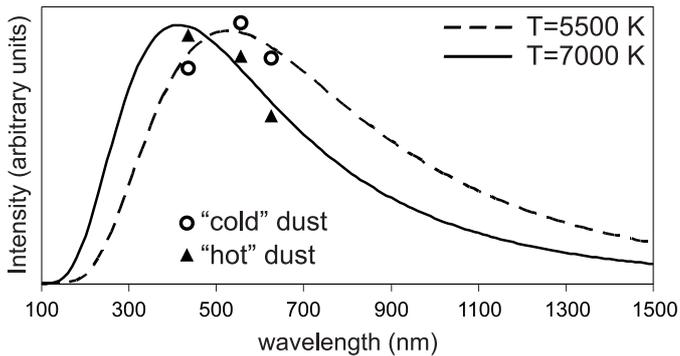}
 \caption{ Triangles: average relative proportions among the intensities in the three HST images
at the "hot" zones of the dust ring (according to fig. ~\ref{thermal} bottom right)
and its fitting to a black body of 7000 K (solid line).
Circles: the same for the "cold" zones of the dust ring, and its fitting to a black
body of 5500 K (dashed line).}
\label{spectra}
\end{figure}

But inside the dust ring, a strange and unexpected feature appears: a narrow zone surrounding the galaxy,
with high equivalent black body temperatures. At first sight, one could think it is an odd
artifact of the temperature estimation method, but indeed it is not. It is in the data:
light coming from this narrow zone of the dust ring is indeed bluer than the one from the rest of the ring.
The average relative intensities in the three HST images in the "hot" dust zone (see fig. ~\ref{spectra}) are higher in the F435W filter, followed by the F555W filter, and are lower in the F625W one, fitting
rather well into a 7000 K black body. Meanwhile in the "cold" dust zones, the intensity peaks in the F555W filter, fitting
better into a 5500 K black body.

Of course this does not indicate the true temperature of the dust, which is in fact much colder (almost an
order of magnitude). But why is bluer the light coming from this region of the dust? The fact is that
infrared Spitzer images of this galaxy \cite{ref9} show this zone of the dust ring
glowing brightly in infrared light, unveiling a disk of stars within the dust ring.
This seems to confirm that the "hot" dust zone is indeed the fingerprint of this stellar formation zone;
some of these stars could be in the foreground of the dust ring, sticking out from under the dust,
unresolved but contributing to the color to the dust.

\begin{figure*}
 \includegraphics[width=\linewidth]{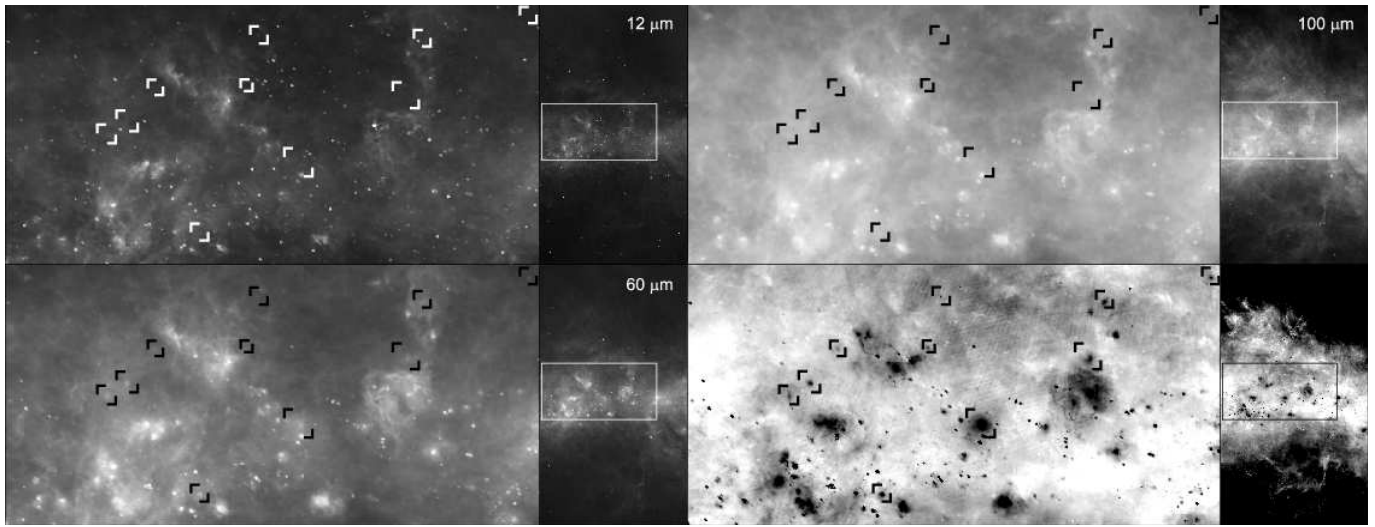}
 \caption{First three panels: infrared images from NASA's IRAS Sky Survey Atlas Galactic Plane Mosaic, centered at 22h, +60$^{\circ}$, respectively for the 12, 60 and 100 Micron bands. Bottom right:
ratio between the real IRAS image at 12 Microns and a synthetic image for the same wavelength generated from the IRAS images
at 60 and 100 Microns, assuming a black body behavior (logarithmic scale). The brighter the image, the bigger the amount of emission from polycyclic aromatic hydrocarbon molecules. Main panels are a zoom of the complete images, shown at right.
Boxes mark the position of some objects appearing clearly in the bottom right image but dim or absent in the three infrared images.} \label{infrared}
\end{figure*}

\section{Synthetic images according to black bodies}
Once one has an estimation of the temperature of a black body from $I_a$ and $I_b$
using eq. (12), it is possible to estimate its emission at any other wavelength, just
using eq. (1) and solving $C$ from the known data:
\begin{equation}
I = I_i \frac{{x^n }}{{x_i^n }}\left( {\frac{{e^{x_i /\tilde T}  - 1}}{{e^{x/\tilde T}  - 1}}} \right)
\end{equation}
where $i$ can be $a$ or $b$. The best results are obtained by choosing the one
closer to the target spectral position $x$. This equation (together with (12)
and (13) to obtain $\tilde T$), gives much better results than eq. (7) and allows to extrapolate
the behavior of black bodies at wavelengths very far away from the input
spectral positions $a$ and $b$.

Equation (15) provides a quick way to identify divergences
from black body behavior or similitudes with it. As an example we will use infrared data from NASA's IRAS Sky Survey Atlas (ISSA). Concretely, images in the 12, 60 and 100 Micron bands from the Galactic Plane Mosaic, centered at 90$^{\circ}$ of
galactic longitude. Figure ~\ref{infrared} first three panels show these IRAS images as insets at right, being the main panels a zoom on their central zone,
centered approx. at Ra 22h, Dec +60$^{\circ}$, with a size of about 30$^{\circ}$x12$^{\circ}$, respectively at wavelengths
12, 60 and 100 Microns.

In the 12 Microns band, the galactic plane is far from a black body behavior, as the emission from polycyclic
aromatic hydrocarbon molecules dominates in the galactic plane at this wavelength, meanwhile in the 60 and 100 Micron bands
dominates the thermal emission from cold dust, following a Planck distribution \cite{ref10}.
 Given that in the 60 and 100 Micron bands we do expect blackbody behavior, we will use the images
at these bands to generate a synthetic image at 12 $\mu$m,
and then divide the real one at 12 $\mu$m by the synthetic one. Hence, the excess of emission in this ratio image will be representative of
the presence of  hydrocarbons. This is what is
shown in the bottom right panel of fig. ~\ref{infrared} (in logarithmic scale, rescaled).

We are not interested in the PAH emission itself but in the cluster of dark objects
which seem behave as black bodies. These objects seem to
be in the foreground, shielding the background hydrocarbon radiation emitted from the Galaxy. Some of them
coincide with nebulae, clearly visible in the three infrared images. But there
are many others (a selection has been marked with boxes) that can hardly be made out in the infrared images,
some nearly imperceptible (but it is clear that they are real objects and not artifacts,
as they show a clear elliptical/circular shape). Comparing with visible images of this region, we see that a part of these objects are
related to stars (standing out the bright one close to the center, whose
location coincides with the supergiant $\lambda$ Cephei), but not all of them appear in optical images
(and vice versa, not all the stars have a counterpart in the "blackbodyness" image).
In all these cases, their equivalent black body
temperature happens to be very low: they are compatible with a black body at $\approx$50 K. What are these objects?
Are these features dust clouds shielding the galactic plane emission?
Does the supergiant $\lambda$ Cephei have an envelope surrounding it, perhaps due to its stellar wind?
Are protostars those globules apparently not associated to a star?
In any case, they seem to have a strong "black body" signature that makes them very evident
under this procedure.

\section{Conclusions}

We have done a new analytical study of Planck's law and we have obtained some new
properties in a formula which was published more than a century ago. The two main outcomes of this analysis
are the deduction of the exact form of the power law lying inside Planck's law, which explains
most of the behavior of the stellar locus in a color-color diagram,
and a new analytical formula to estimate black body temperatures from the ratio of
intensities measured at two different spectral positions, which also provides a
method to estimate black body emissions in any other spectral position.

As a result, we have introduced a new tool for data analysis.
In this paper we have shown how it can be used to check if a data reduction has been
done properly, to characterize the effective wavelengths of filters, and how it
allows to obtain very fast
thermal images of effective black body temperatures, which can give information about "hot"
zones, as in the case of the stellar formation region in the dust ring of the Sombrero Galaxy.
It allows to generate also very fast synthetic images at any other wavelength following a black body
behavior. We have shown how this can be useful to identify interesting objects, as in the IRAS data analyzed in the previous section.

Other possible applications will be object of future consideration, as for example to compare if the method of
black body temperature estimation can be useful or competitive in remote sensing, for land-surface temperature
measurements, comparing it with methods as the Split-Window method (\cite{ref11}, which calculates the temperature
from flux measurements in two infrared bands).

Finally, this results are going to be implemented as a set of tools
in the astronomical image processing software PixInsight (http://pixinsight.com).
\acknowledgments
I would like to thank Vicent Peris for his invaluable help in the data processing and useful
brainstormings. I would like also to thank Eusebio Llacer for his help with the English, and
Vicent Martinez, Alberto Fernandez, Bartolo Luque and Juan Fabregat for their useful suggestions and comments.
I want to acknowledge support from MICINN through CONSOLIDER projects AYA2006-14056
and CSD2007-00060, and project AYA2010-22111-C03-02. All the images for this paper have been processed using
PixInsight.

\end{document}